\documentclass[oneside,english]{amsart}
\usepackage[T1]{fontenc}
\usepackage[latin9]{inputenc}
\usepackage{textcomp}
\usepackage{amstext}
\usepackage{amsthm}
\usepackage{amssymb}

\makeatletter
\numberwithin{equation}{section}
\numberwithin{figure}{section}

\usepackage{babel}

\usepackage{babel}

\usepackage{babel}

\usepackage{babel}

\makeatother

\usepackage{babel}
\begin{document}

\title{Relative Scales of the GUT and Twin Sectors in an F-theory model}

\author{C. Herbert Clemens and Stuart Raby}

\date{December 12, 2021}

\email{clemens.43@osu.edu, raby.1@osu.edu}
\begin{abstract}
In this letter we analyze the relative scales for the GUT and twin
sectors in the F-theory model discussed in Ref. \cite{Clemens-3}.
There are a number of volume moduli in the model. The volume of the
GUT surface in the visible sector {[}sector(1){]} (with the Wilson
line GUT breaking) defines the GUT scale $M_{G}\sim3\times10^{16}~GeV$
as the unification scale with precise gauge coupling unification of
$SU(3)\times SU(2)\times U(1)_{Y}$. We choose the GUT coupling constant,
$\alpha_{G}^{-1}\sim24$. We are then free to choose the ratio $\alpha_{G}(2)/\alpha_{G}(1)=m_{1}/m_{2}$
with $m_{1}$ and $m_{2}$ independent volume moduli associated with
the directions perpendicular to the two asymptotic GUT surfaces. We
then analyze the effective field theory of the twin sector (2), which
may lead to a SUSY breaking gaugino condensate. Of course, all these
results are subject to the self-consistent stabilization of the moduli.
\end{abstract}

\maketitle

\section{Relative Scales in F-theory GUT}

The effective low energy field theory of an $F$-theory $\mathrm{GUT}$
is defined on a real $10$-dimensional manifold $\mathbb{M}^{10}=\mathbb{R}^{3,1}\times B_{3}$
where $B_{3}$ is a smooth complex projective complex $3$-fold with
ample anti-canonical bundle, that is, a Fano threefold. Gravity fills
all of this real $10$-dimensional space-time while the GUT theory
resides on a 7-brane, $S_{\mathrm{GUT}}\times \mathbb{R}^{3,1}$, with $S_{\mathrm{GUT}}$ 
given by a smooth two-dimensional anti-canonical complex surface
$S_{\mathrm{GUT}}\subseteq B_{3}$. The $\mathrm{GUT}$ surface $S_{\mathrm{GUT}}$
is defined by the vanishing of $z\in H^{0}\left(K_{B_{3}}^{-1}\right)$.

\subsection{Semi-stable degeneration of the $F$-theory model}

More precisely, as in \cite{Clemens-2} we consider the product
\[
B_{3}=\mathbb{P}_{\left[u_{0},v_{0}\right]}\times B_{2}
\]
$B_{2}$ is a fixed del Pezzo surface with $\mathbb{Z}_{4}$-symmetry.
For the affine coordinates
\[
\left(a,b\right)\in\mathbb{C}^{2}\subseteq\mathbb{P}_{a}^{1}\times\mathbb{P}_{b}^{1}
\]
we set
\[
\begin{array}{c}
a=\delta^{1/2}\frac{u_{0}-v_{0}}{u_{0}+v_{0}}\\
b=\delta^{1/2}\frac{u_{0}+v_{0}}{u_{0}-v_{0}}
\end{array}
\]
and consider the closure $\mathbb{P}_{\left[\delta\right]}$ of the
subset
\[
a\text{·}b=\delta\in\left[0,1\right]
\]
so that
\[
B_{3,\delta}=\mathbb{P}_{\left[\delta\right]}\times B_{2}
\]
and for $\delta\neq0$
\[
B_{3,\delta}\cong B_{3}.
\]

Here $B_{2}$ is a divisor in $D_{6}\times\mathbb{P}^{1} \equiv D_{6}\times\mathbb{P}_{\left[k,l\right]}$
endowed with its canonical toric metric $g_{\left(6\right)}$ as in
(1.3) of \cite{Guillemin} with respect to which the $\mathbb{Z}_{4}$-action
$T_{0}$ as in Section 6.1 of \cite{Clemens-2} is isometric with
finite fixpoint set.\footnote{This metric is not the Kähler-Einstein metric on $D_{6}$. See \cite{Doran}.}
We recall that $D_{6}$ is the blow-up of $\mathbb{P}^{2}=\mathbb{P}_{\left[a,b,c\right]}$ at the three
points where two of its coordinates vanish.  To obtain $B_2$ we pick one general point in $D_{6}$ and blow up the
four points of its orbit under the $\mathbb{Z}_{4}$ symmetry generated by $T_{0}$.

To do this we proceed as follows.  Any four points of $\mathbb{P}_{\left[a,b,c\right]}$, no three of
which are collinear, can be written as the common zeros of two homogeneous
quadratic forms, $q_{1}\left(a,b,c\right)$ and $q_{2}\left(a,b,c\right)$.
Since the intersection set is invariant under the action of $T_{0}$,
the $q_{j}$ can also be chosen to be invariant. Then
\[
B_{2}\subseteq D_{6}\times\mathbb{P}^{1} \equiv D_{6}\times\mathbb{P}_{\left[k,l\right]}
\]
is the set
\[
\left\{ \left|\begin{array}{cc}
q_{1}\left(a,b,c\right) & q_{2}\left(a,b,c\right)\\
k & l
\end{array}\right|=0\right\}
\] that defines the blow up as a divisor in $D_{6}\times\mathbb{P}_{\left[k,l\right]}$
by smoothly inserting a copy of $\mathbb{P}_{\left[k,l\right]}$ at each the four common zeros of $q_{1}\left(a,b,c\right)$ and $q_{2}\left(a,b,c\right)$ in $D_{6}$. We can define the metric $g_{\left(2\right)}$ on $B_{2}$
as the one induced from the product metric on $D_{6}\times\mathbb{P}_{\left[k,l\right]}$
by restriction. Thus we have one additional real degree of freedom
in the choice of the scaling constant on the standard $SU\left(2\right)$-invariant
metric on $\mathbb{P}^{1}$. Summarizing, the metric induced on
\[
B_{3,\delta}=B_{2}\times\mathbb{P}_{\left[\delta\right]}\subseteq\left(D_{6}\times\mathbb{P}_{\left[k,l\right]}\right)\times\left(\mathbb{P}_{a}^{1}\times\mathbb{P}_{b}^{1}\right)
\]
and all its submanifolds is the metric induced by inclusion from the
product of the four given metrics on $D_{6}\times\mathbb{P}_{\left[k,l\right]}\times\mathbb{P}_{a}^{1}\times\mathbb{P}_{b}^{1}$.
The canonical metric on $D_{6}$ is the standard metric given by its
toric structure \cite{Guillemin} and the metric on each of the three
$\mathbb{P}^{1}$'s is a positive real multiple of the standard $SU\left(2\right)$-invariant
metric. We let
\[
\begin{array}{c}
m_{1}:=Vol\left(\mathbb{P}_{a}^{1}\right)\\
m_{2}:=Vol\left(\mathbb{P}_{b}^{1}\right).
\end{array}
\]
This allows two additional scaling constants, the first giving volume
$m_{1}$ to the standard $SU\left(2\right)$-invariant metric on $\mathbb{P}_{a}$
and the second giving volume $m_{2}$ to the standard $SU\left(2\right)$-invariant
metric on $\mathbb{P}_{b}$.

The Einstein-Hilbert action is given by
\begin{equation}
S_{EH}\sim M_{\ast}^{8}\int_{\mathbb{R}^{3,1}\times B_{3,\delta}}R\sqrt{-g_{\delta}}d^{10}x.\label{eq:1.1}
\end{equation}
As a consequence, the four-dimensional Planck constant is given by
\begin{equation}
M_{Pl}^{2}\simeq M_{\ast}^{8}\text{·}Vol\left(B_{3,\delta}\right).\label{eq:1.2}
\end{equation}

The semi-stable limit of the $F$-theory geometry as $\delta$ goes
to zero is the union of two components or 'gauge sectors'
\[
B_{3}^{\left(1\right)}\cup B_{3}^{\left(2\right)}
\]
crossing along a copy of $B_{2}$ over $\left(a,b\right)=\left(0,0\right)$
where

\[
\begin{array}{c}
B_{3}^{\left(1\right)}=\mathbb{P}_{a}^{1}\times B_{2,0}\\
B_{3}^{\left(2\right)}=\mathbb{P}_{b}^{1}\times B_{2,0}.
\end{array}
\]
We call $B_{3}^{\left(1\right)}$ with induced metric $g_{1}$ the
'visible sector' and $B_{3}^{\left(2\right)}$ with induced metric
$g_{2}$ the 'hidden or twin sector.' Thus
\[
Vol\left(B_{3,0}\right)=Vol\left(B_{2}\right)\text{·}\left(m_{1}+m_{2}\right).
\]

\subsection{Asymptotic position of $S_{\mathrm{GUT}}$}

As $\delta$ varies, the $\mathrm{GUT}$ surface $S_{\mathrm{GUT},\delta}$
is defined by the vanishing of
\[
z_{\delta}=\delta\text{·}z+\left(1-\delta\right)\text{·}(u_{0}^{2}-v_{0}^{2})\text{·}n_{0}\in H^{0}\left(K_{B_{3,\delta}}^{-1}\right)
\]
where $n_{0}\in H^{0}\left(K_{B_{2,0}}^{-1}\right)$ is a section
in the $\left(-1\right)$-eigenspace for the $\mathbb{Z}_{4}$-action
on $H^{0}\left(K_{B_{2}}^{-1}\right)$. We let $C\subseteq B_{2}$
denote the arithmetic genus-one curve on $B_{2}$ defined by $\left\{ n_{0}=0\right\} \subseteq B_{2}$.

Here the gauge action is given by
\[
S_{gauge}\sim-M_{\ast}^{4}\int_{\mathbb{R}^{3,1}\times B_{3,0}}\left(Tr(F_{1}^{2})\sqrt{-g_{1}}+Tr(F_{2}^{2})\sqrt{-g_{2}})\right)\delta^{2}(z_{0})~d^{10}x
\]
where $F_{i}$ denotes the (limiting) curvature tensor of the Yang-Mills
connection on the $i$-th gauge sector $B_{3}^{\left(i\right)}$ and
$\delta^{2}(z_{0})$ is the standard distribution supported on
\[
\left\{ z_{0}=0\right\} =S_{\mathrm{GUT},0}=S_{1}\cup S_{2}
\]
where
\[
\begin{array}{c}
S_{1}:=\left(\left\{ a=\infty\right\} \times B_{2}\right)\cup\left(\mathbb{P}_{a}^{1}\times C\right)\\
S_{2}:=\left(\left\{ b=\infty\right\} \times B_{2}\right)\cup\left(\mathbb{P}_{b}^{1}\times C\right).
\end{array}
\]
Therefore
\[
Vol\left(S_{i}\right)=M_{G}\left(i\right)^{-4}=Vol\left(B_{2}\right)+m_{i}\text{·}Vol\left(C\right).
\]
(See Appendix \ref{sec:Asymptotic-position-of} for more detail on
the decomposition $S_{1}\cup S_{2}$ of $S_{\mathrm{GUT},0}$.)

\section{Scaling the effective 4-D theory}

Hence in the effective $4$-dimensional theory we should have $\mathrm{GUT}$
coupling constant
\[
\alpha_{G}(i)^{-1}\sim M_{\ast}^{4~}Vol(S_{i}),~~i=1,2.
\]
Substituting we obtain
\[
\alpha_{G}(i)^{-1}\sim M_{\ast}^{4~}\left(Vol(B_{2})~+~m_{i}~Vol\left(C\cup C^{opp}\right)\right),~~i=1,2.
\]

We then find
\[
\begin{array}{c}
\alpha_{G}\left(i\right)M_{Pl}\sim\frac{\left[Vol\left(B_{2}\right)\left(m_{1}+m_{2}\right)\right]^{1/2}}{Vol(B_{2})+m_{i}\text{·}Vol\left(C\cup C^{opp}\right)\text{·}\int_{B_{2,0}}\left|n_{0}\right|^{2}}\\
=\frac{\left[Vol\left(B_{2}\right)(m_{1}+m_{2})\right]^{1/2}}{Vol(B_{2})\left(1+K\text{·}m_{i}\right)}.
\end{array}
\]
Therefore the relative size of the GUT coupling constants and the
GUT scales for the visible and twin sectors depends on the relative
sizes of the perpendicular directions in $B_{3,0}$ to $B_{2,}$.

Let's define the sector labeled $(1)$ as the visible sector with
GUT coupling constant, $\alpha_{G}(1)^{-1}=24$ at the GUT scale $M_{G}(1)=3\times10^{16}~GeV$.
Then the twin sector is sector $(2)$. The ratio $\frac{\alpha_{G}(2)}{\alpha_{G}(1)}=\frac{1+Km_{1}}{1+Km_{2}}=\left(\frac{M_{G}(2)}{M_{G}(1)}\right)^{4}$.
Let's take $\alpha_{G}(2)^{-1}=8.7$ or $\frac{1+Km_{1}}{1+Km_{2}}=2.8$
and $M_{G}(2)=3.9\times10^{16}~GeV$. Below the scale $M_{G}(2)$
the effective field theory is $SU(3)\times SU(2)\times U(1)_{Y}$,
just as in the visible sector. However the twin QCD coupling will
become strong at a scale much greater than the visible QCD scale.
The effective twin QCD theory has $N_{C}=3$ and $N_{flavors}=6$.
Hence it is described by Seiberg duality \cite{Seiberg:1994pq}. In
the magnetic phase, we have an effective superpotential given by
\begin{equation}
W=q^{ia}{T_{i,a}}^{j,b}\bar{q}_{j,b}+\lambda_{ij}^{u}q^{ia}{H_{u}}_{a}\bar{q}_{j1}+\lambda_{ij}^{d}q^{ia}{H_{d}}_{a}\bar{q}_{j2}
\end{equation}
where $q$ ($\bar{q}$) are left-handed color triplets (anti-triplets)
with the family index, $i=1,2,3$, and $SU(2)_{isospin}$ index, $a=1,2$.
When $\langle q\rangle_{0}=\langle\bar{q}\rangle_{0}=0$, the theory
has a flat direction for the fundamental meson field, ${T_{i,a}}^{j,b}$.
Note, since the twin electroweak group is gauged, we should identify
${T_{i,a}}^{j,1}\equiv{({T_{u}}_{a})_{i}}^{j}$ and ${T_{i,a}}^{j,2}\equiv{({T_{d}}_{a})_{i}}^{j}$.
The twin supersymmetric $SU(2)\times U(1)_{Y}$ gauge interactions
introduce a quartic potential for $T_{u},~T_{d},~H_{u},~H_{d}$ such
that there is a flat direction for $\langle{({T_{u}}_{1})_{i}}^{j}\rangle=\langle{({T_{d}}_{2})_{i}}^{j}\rangle=T{\delta_{i}}^{j}$
and ${H_{u}}_{1}={H_{d}}_{2}=T$. Then all twin quarks and leptons
obtain mass at the scale $T$ and, moreover, the twin electroweak
gauge symmetry is broken down to twin $U(1)_{EM}$. For $T\sim M_{G}(2)$,
we find a twin gluino condensate occurs at the scale $\Lambda_{tQCD}=T\exp(-\frac{2\pi}{9\alpha_{G}(2)})\sim9\times10^{13}~GeV$.

We expect that the effective 4D QCD Lagrangian contains a term
\begin{equation}
L\supset\frac{1}{2}\int d^{2}\theta\sum_{i=1}^{2}\left(\frac{S(i)}{4}TrW^{\alpha}W_{\alpha}(i)+h.c.\right)
\end{equation}
with
\begin{equation}
S(i)=\frac{1}{4\pi\alpha_{G}(i)}+i\theta=e^{\ln(K(i)m_{i})-\phi}+ib,
\end{equation}
where $\phi,b$ is the dilaton, axion fields, $m_{i}$ is as above
and $K(i)\sim M_{\ast}^{6~}M_{G}(i)^{-4}$ \cite{Nilles:2004zg,Gorlich:2004qm,Dundee:2010sb}.
As a consequence, the twin QCD condensate will contribute SUSY breaking
effects to both the twin and visible sectors of the theory. In this
local SUSY theory, we find an effective low energy SUSY breaking scale
given by
\begin{equation}
m_{3/2}=\Lambda_{tQCD}^{3}/M_{Pl}^{2}\sim130~TeV.
\end{equation}
Of course, whether supersymmetry is broken (or not) depends on stabilizing
all the moduli.

The low energy supersymmetric theory contains 3 families of twin neutrino
superfields (assuming that the three right-handed neutrinos obtain
mass near the GUT scale), 19 chiral charged and neutral Higgs superfields
(which include the massless components of $H_{u},H_{d}$ and $T_{u},T_{d}$),
and the twin photon superfield. Renormalizing from $M_{G}(2)$ we
find $\alpha^{tEM}(m_{3/2})\sim1/105$. There do not appear to be
any portals to the twin sector. Clearly, the twin sector introduces
new candidates for dark matter, but a complete analysis of the cosmological
implications of this sector for the theory is beyond the scope of
the present paper.

\section{Conclusion}

In conclusion, we have analyzed the relative scales of the visible
and twin sectors in the global $F$-theory GUT with Wilson line breaking
given in \cite{Clemens-3}. We have found that there is sufficient
freedom to have independent GUT scales and couplings in order to have
interesting physics coming from the twin sector. In particular, if
we assume that the GUT coupling for the twin sector is larger than
that of the visible sector, then it is possible to spontaneously break
the twin electroweak theory at the GUT scale with all twin quarks
and charged leptons obtaining mass at that scale. In addition, a twin
gluino condensate can then occur at a scale of order $\Lambda_{tQCD}\sim9\times10^{13}~GeV$
which leads to an effective low energy SUSY breaking scale, $m_{3/2}=\Lambda_{tQCD}^{3}/M_{Pl}^{2}\sim130~TeV$
which affects both the twin and visible sectors. There is clearly
more analysis that needs to be done on the consequences of these results,
including the stabilization of moduli, which we leave for the future.
For example, the model also includes $11~D_{3}$ branes and fluxes
which must be considered \cite{Cecotti:2009zf}.

\appendix

\section{\label{sec:Asymptotic-position-of}Asymptotic position of $S_{\mathrm{GUT}}$
with regards to the Heterotic theory }

The Heterotic dual is the smooth Calabi-Yau threefold $\left[V_{3}=B_{3}^{\left(1\right)}\cap B_{3}^{\left(2\right)}\right]$
over $a=b=0$. Each $B_{3}^{\left(i\right)}$ encodes the structure
of an $E_{8}$-bundle with Yang-Mills connection on the Heterotic
model $V_{3}$ via the equivalence of Yang-Mills $E_{8}$-bundles
and $dP_{9}$-bundles given by the dictionary in §4.2 of \cite{Friedman:1997yq}.
A slightly subtle point in the encoding is that the factor $n_{0}$
in $z_{0}$ is irrelevant to the determination of the $E_{8}$-bundle,
since over each point $b_{2}\in B_{2,0}$, $n_{0}\left(b_{2}\right)$
simply re-scales the vector $\left[s,t\right]$ in the Weierstrass
form $\left[y^{2}=4x^{3}-\left(g_{2}t^{4}-\beta_{1}st^{3}-\ldots-\beta_{4}s^{4}\right)x-\left(g_{3}t^{6}-\alpha_{2}s^{2}t^{4}-\ldots-\alpha_{6}s^{6}\right)\right]$
for the equation of each $dP_{9}$-fiber of $B_{3}^{\left(i\right)}/B_{2}$.
Thus the Weierstrass form, and so the isomorphism class of the $dP_{9}$-bundle,
is unaffected. Said otherwise, Heterotic ``$S_{\mathrm{GUT}}$\textquotedbl{}
is simply the union of the section $\{a=\infty\}$ of $B_{3}^{\left(1\right)}/B_{2}$
and the section $\{b=\infty\}$ of $B_{3}^{\left(2\right)}/B_{2}$
.

\end{document}